\documentclass[preprint,aps,nofootinbib]{revtex4}
\usepackage{graphicx}
\usepackage{amsmath}
\usepackage{amsfonts,amsbsy}
\usepackage{amssymb}

\newcommand{\rv}[1]{ {\bf r}_{#1} }
\newcommand{\ddd}{,\ldots,}
\newcommand{\kb}{k_B}
\def\lsim{\mathrel{\rlap{\lower4pt\hbox{\hskip1pt$\sim$}}
    \raise1pt\hbox{$<$}}}               
\def\gsim{\mathrel{\rlap{\lower4pt\hbox{\hskip1pt$\sim$}}
    \raise1pt\hbox{$>$}}}  

\def\sig{\sigma_{\textrm{tr}}\left(\gamma\right)}

\def\sigbar{\overline{\sigma}_{\textrm{tr}}}

\begin{document}

\title{Quasiclassical molecular dynamics for the dilute Fermi gas
at unitarity}

\author{K.~Dusling and T.~Sch\"afer}

\affiliation{Department of Physics, North Carolina State University,
Raleigh, NC 27695}

\begin{abstract}
We study the dilute Fermi gas at unitarity using molecular 
dynamics with an effective quantum potential constructed 
to reproduce the quantum two-body density matrix at unitarity.  
Results for the equation of state, the pair correlation function 
and the shear viscosity are presented. These quantities are 
well understood in the dilute, high temperature, limit. Using
molecular dynamics we determine higher order corrections in 
the diluteness parameter $n\lambda^3$, where $n$ is the density
and $\lambda$ is the thermal de Broglie wave length. In the 
case of the contact density, which parameterizes the short 
distance behavior of the correlation function, we find that the results 
of molecular dynamics interpolates between the truncated
second and third order virial expansion, and are in excellent 
agreement with existing T-matrix calculations.  For the shear 
viscosity we reproduce the expected scaling behavior at high
temperature, $\eta\sim 1/\lambda^3$, and we determine the leading 
density dependent correction to this result. 
\end{abstract}
\maketitle 

%%%%%%%%%%%%%%%%%%%%%%%%%%%%%%%%%%%%%%%%%%%%%%%%%%%%%%%%%%%%%%%%%%%%%%%%%
\section{Introduction}
%%%%%%%%%%%%%%%%%%%%%%%%%%%%%%%%%%%%%%%%%%%%%%%%%%%%%%%%%%%%%%%%%%%%%%%%%

 There has been a significant amount of interest in equilibrium
and transport properties of cold atomic quantum gases near a Feshbach 
resonance \cite{bloch2008,giorgini2008,Schafer:2009dj,Adams:2012th}.
The Feshbach resonance is used to tune the energy of a molecular bound 
state to the threshold of a scattering state, causing the scattering 
length to diverge.  At resonance the scattering cross-section 
is universal, independent of the microscopic details, and is limited 
only by unitarity.  In this regime there are no small parameters that 
can be used to justify a perturbative expansion.

The only ab-initio approach to the problem is the quantum Monte-Carlo (QMC)
method \cite{Carlson:2003zz,Astrakharchik:2004zz,Bulgac:2005pj,Lee:2005it,goulko2010,Drut:2011tf,Endres:2012cw}.
QMC simulations have been used successfully to compute thermodynamic 
properties, but it is very difficult to use imaginary time QMC 
simulations to compute real-time response functions and transport 
coefficients (see, however, \cite{Wlazlowski:2012jb} for a recent attempt). 
This implies that we lack reliable bench mark calculations for 
experimental attempts to determine the viscosity of a degenerate Fermi gas near 
unitarity.  While the viscosity can be computed reliably at both high 
\cite{Massignan:2005zz,Bruun:2006kj,Bruun:2005en} and low 
\cite{Rupak:2007vp,Mannarelli:2012su} temperature there is no systematic 
approach in the strongly coupled regime $T\sim T_F$, where $T_F$ is the 
Fermi temperature. A number of authors have used diagrammatic methods 
to study transport properties in this regime \cite{Enss:2010qh,Guo:2011}, 
but it is not a priori clear what kind of diagrams have to be included. 

In this work we introduce a novel approach to the dynamics of quantum 
gases. The method is based on a classical molecular dynamics simulation 
in which quantum effects are encoded in an effective classical interaction 
among the atoms. The quasi-classical $N$-body interaction is constructed 
such that the classical calculation exactly reproduces the diagonal
component of the quantum  $N$-body density matrix. This implies, in 
particular, that the $N$'th quantum virial coefficient is reproduced. 
In this paper we restrict ourselves to two-body terms in the 
interaction. In this case the second virial coefficient will be 
reproduced exactly, and through molecular dynamic simulations, the 
one and two-body components of the higher virial coefficients are 
resumed. The strength of the molecular dynamics method is that very 
complicated many body correlations are taken into account. The drawback 
is that genuine quantum many-body effects such as pairing and superfluidity 
are not included.  Quasi-classical molecular dynamics has been used 
successfully in the study of strongly correlated Coulomb plasmas 
\cite{Golubnychiy:2001,Bonitz:2003}.  However, our work differs in
that the quasi-classical potential for unitary fermions is of purely 
quantum mechanical origin.  In contrast, the quasi-classical Coulomb 
potential, known as the Kelbg potential \cite{Kelbg:1963}, is a small 
correction to the classical $1/r$ behavior. Quasi-classical methods
were also used by Feynman and Kleinert to study the high temperature 
limit of the partition function for simple quantum systems
\cite{Feynman:1986ey}.

 This paper is organized as follows. In Sect.~\ref{sec:model} and 
\ref{sec:veff} we introduce the method and derive the quasi-classical 
potential at unitarity. In Sect.~\ref{sec:md} we describe the molecular 
dynamics simulations and in Sect.~\ref{sec:results} we present results 
for the equation of state, the pair correlation function, and the shear 
viscosity.

%%%%%%%%%%%%%%%%%%%%%%%%%%%%%%%%%%%%%%%%%%%%%%%%%%%%%%%%%%%%%%%%%%%%%%%%%
\section{The partition function}
\label{sec:model}
%%%%%%%%%%%%%%%%%%%%%%%%%%%%%%%%%%%%%%%%%%%%%%%%%%%%%%%%%%%%%%%%%%%%%%%%%

 In this section we introduce a classical partition function which 
is equivalent, order-by-order in a cluster expansion, to the full 
quantum partition function. The classical partition function 
depends on $k$-body potentials which can be determined from 
the quantum mechanical Slater sums. In this work we will restrict
ourselves to the case $k=2$, which means that the second virial 
coefficient is reproduced exactly. We will follow the notation 
used in \cite{huang1987statistical}.

 The quantum mechanical partition function for a system of 
$N$ particles is given by 
\begin{align}
\mathcal{Z}_N=\frac{1}{N!\lambda^{3N}}\int\left(d\rv{1} 
 \ldots d\rv{N}\right)W^{(N)}\left(\rv{1}\ddd\rv{N}\right)\, , 
\end{align}
where we have defined the $N$-particle Slater sums 
\begin{align}
W^{(N)}\left(\rv{1}\ddd\rv{N}\right)
   = N!\lambda^{3N}\sum_\alpha\vert \Psi_\alpha
   \left(\rv{1}\ddd\rv{N}\right)\vert^2 e^{-\beta E_\alpha}\, . 
\end{align}
In this  expression $\lambda = \sqrt{\frac{2\pi\hbar^2}{m\kb T}}$ 
is the thermal wavelength, $\beta=1/T$ is the inverse 
temperature, and $\Psi_\alpha(\rv{1}\ddd\rv{N})$ is
the wave function of an $N$ particle state with energy $E_\alpha$. 
The partition function can be expanded systematically in powers 
of $n\lambda^3$ in terms of the virial coefficients $b_l$,
\begin{align}
\mathcal{Z}_N
  =  \sum_{\{m_l\}}\prod_{l=1}^N
    \frac{1}{m_l!}\left(\frac{V}{\lambda^3} b_l\right)^{m_l}\, ,
\label{eq:qvirial}
\end{align}
where $\{m_l\}$ is a set of integers $m_l\geq 0$ that satisfies 
the constraint $\sum_{l=1}^N lm_l=N$. The corresponding expression 
for the pressure has the form
\begin{align}
\frac{P}{kT}=\frac{1}{\lambda^3}\sum_{l=1}^\infty b_l z^l \, , 
\end{align}
where $z=e^{\beta\mu}$ is the fugacity and $\mu$ is the chemical 
potential. Each virial coefficient $b_N$ can be expressed 
in terms of integrals over the $N$-particle Slater sums.  For example
\begin{align}
b_2 &= \frac{1}{2! \lambda^3 V}\int d\rv{1} d\rv{2} 
   \left[W^{(2)}(\rv{1},\rv{2})-W^{(1)}(\rv{1})W^{(1)}(\rv{2})\right]\;, \\
b_3 &= \frac{1}{3! \lambda^6V}\int d\rv{1} d\rv{2} d\rv{3}
   \left[W^{(3)}(\rv{1},\rv{2},\rv{3})-W^{(2)}(\rv{1},\rv{2})
         W^{(1)}(\rv{3})\right.\nonumber\\
   & -\left.W^{(2)}(\rv{2},\rv{3})W^{(1)}(\rv{1})-W^{(2)}(\rv{3},\rv{1})
    W^{(1)}(\rv{2})+2W^{(1)}(\rv{1})W^{(1)}(\rv{2})W^{(1)}(\rv{3})\right]\, .
\end{align}
We may compare these results to the corresponding expressions for a 
classical system. We consider the most general partition function 
containing arbitrary $N$-body interactions
\begin{align}
\mathcal{Z}_N =\frac{1}{N!\lambda^{3N}}
  \int\left(d\rv{1} \ddd d\rv{N}\right)
     e^{-\beta\sum_{i< j} v_{ij} -\beta\sum_{i< j< k} v_{ijk} + \cdots } \; .
\label{eq:Qcl}
\end{align}
The virial expansion of the classical partition function has the 
same form as the quantum expansion in Eq.~(\ref{eq:qvirial}), but 
the expressions for the virial coefficients are different. We have
\begin{align}
b_2 &= \frac{1}{2! \lambda^3 V}\int d\rv{1} d\rv{2} 
   \left[e^{-\beta v_{12}}-1\right] \;,\\
b_3 &=\frac{1}{3! \lambda^6V}\int d\rv{1} d\rv{2} d\rv{3} 
   \left[e^{-\beta \left(v_{123}+v_{12}+v_{23}+v_{13}\right)}
   -e^{-\beta v_{12}}-e^{-\beta v_{23}}-e^{-\beta v_{13}}+2\right]\; .
\end{align}
It is clear that one can construct a classical $N$-body potential
so that the classical and quantum virial coefficients agree order 
by order.  For example, we can construct effective 2 and 3-body 
potentials
\begin{align}
\label{eq:v2}
v_{ij}  &= -\beta^{-1}\log\left(W^{(2)}\left(\rv{i},\rv{j}\right)\right)\, , \\
\label{eq:v3}
v_{ijk} &= -\beta^{-1}\log\frac{W^{(3)}(\rv{1},  \rv{2},\rv{3})}
     {W^{(2)}(\rv{1},\rv{2})W^{(2)}(\rv{2},\rv{3})W^{(2)}(\rv{3},\rv{1})}\, , 
\end{align} 
such that the classical system described by the partition function in
Eq.~(\ref{eq:Qcl}) will have the same virial coefficients as the quantum 
mechanical system at the same order. In practice we will truncate
the expansion at second order. We note that even in this case we
retain all contributions to the third and higher virial coefficients 
that arise from powers of $W^{(2)}$. Only genuine three and 
higher-body correlations are missing.

%%%%%%%%%%%%%%%%%%%%%%%%%%%%%%%%%%%%%%%%%%%%%%%%%%%%%%%%%%%%%%%%%%%%%%%%%
\section{Quasi-classical two-body potential at unitarity}
\label{sec:veff}
%%%%%%%%%%%%%%%%%%%%%%%%%%%%%%%%%%%%%%%%%%%%%%%%%%%%%%%%%%%%%%%%%%%%%%%%%

As discussed in the previous section the first non-trivial virial 
coefficient can be reproduced by introducing the effective two-body 
potential defined in Eq.~(\ref{eq:v2}). The potential depends on
the logarithm of the two-particle Slater sum $W^{(2)}$. This quantity
is defined in terms of the solutions of the two-particle Schr\"odinger 
equation
\begin{align}
  &\mathcal{H}\Psi_\alpha\left(\rv{1},\rv{2}\right) 
= E_\alpha\Psi_\alpha\left(\rv{1},\rv{2}\right)\, ,  \nonumber\\
  &\mathcal{H} = -\frac{\hbar^2}{2m}\left({\bf \nabla}_{r_1}^2 
 + {\bf \nabla}_{r_2}^2\right) + V\left(\vert \rv{1}-\rv{2}\vert \right)\; . 
\end{align}
In the case of a spherically symmetric interaction the two 
particle Slater sum is only a function of the relative coordinate
$r=\vert\rv{2}-\rv{1}\vert$. We find
\begin{align}
W^{(2)}\left(\rv{}\right)= 2^{5/2}\lambda^{3}\sum_{l}\frac{(2l+1)}{4\pi}
\sum_{k} \;  R_{kl}(r)^2 e^{-\beta\epsilon_k}\; ,
\end{align}
where $R_{kl}(r)$ satisfies the radial Schr\"odinger equation
\begin{align}
\left[\frac{\partial^2}{\partial r^2} 
  + \frac{2}{r}\frac{\partial}{\partial r}
  + \left(k^2-\frac{l(l+1)}{r^2}-V(r)\right)\right]R_{kl}(r)=0,
 \;\;\;\;\;\;k^2\equiv \frac{m}{\hbar^2}\epsilon_k \; .
\end{align}

%%%%%%%%%%%%%%%%%%%%%%%%%%%%%%%%%%%%%%%%%%%%%%%%%%%%%%%%%%%%%%%%%%%%%%%%%
\subsection{Free particle}
\label{sec:v_free}
%%%%%%%%%%%%%%%%%%%%%%%%%%%%%%%%%%%%%%%%%%%%%%%%%%%%%%%%%%%%%%%%%%%%%%%%%

 It is instructive to begin by deriving the effective classical potential 
for a free particle. This potential takes into account the effects of 
quantum statistics: repulsion for identical fermions, or attraction for 
identical bosons. The wave function of a free particle is
\begin{align}
R_{kl}(r) = \sqrt{2k^2}j_l(kr)\; ,
\end{align}
and the corresponding two-particle Slater sum is
\begin{align}
W^{(2)}\left(\rv{}\right) = 2^{7/2}\lambda^{3}
 \sum_{l}\frac{(2l+1)}{4\pi^2} \int k^2 dk\;  
   j_l(kr)^2  e^{-\frac{\lambda^2 k^2}{2\pi}}\;.
\end{align}
We compute the Slater sum for identical bosons and fermions by 
restricting the sum over all states to  $l=\textrm{even}$ 
or $l=\textrm{odd}$, respectively. We find
\begin{align}
W^{(2)} \left(\rv{}\right) = 
   \frac{2^{1/2}\lambda^{3}}{\pi^2} \int dk\; k^2\;
   \left[1\pm \textrm{sinc}(2kr)\right] e^{-\frac{\lambda^2 k^2}{2\pi}} 
 = 1 \pm e^{-\frac{2\pi r^2}{\lambda^2}}\; , 
\end{align}
where $\textrm{sinc}(x)=\textrm{sin}(x)/x$ and we have made use of 
the identities
\begin{align}
\sum_{l=0,2,4,...}(2l+1) j^2_{l}(kr) &= \frac{1+\textrm{sinc}(2kr)}{2} \, ,
   \nonumber\\ 
\sum_{l=1,3,5,...}(2l+1) j^2_{l}(kr) &= \frac{1-\textrm{sinc}(2kr)}{2} \, .
\label{eq:jlid}
\end{align}
The resulting potentials are therefore
\begin{align}
\label{u_id}
 u_{eff}^{\textrm{ideal}} = -\kb T
  \log\left(1\pm e^{-2\pi r^2/\lambda^2}\right)\, ,
\end{align}
with the `+' sign for bosons and the `-' sign for fermions.

%%%%%%%%%%%%%%%%%%%%%%%%%%%%%%%%%%%%%%%%%%%%%%%%%%%%%%%%%%%%%%%%%%%%%%%%%
\subsection{Classical potential at unitarity}
\label{sec:v_unit}
%%%%%%%%%%%%%%%%%%%%%%%%%%%%%%%%%%%%%%%%%%%%%%%%%%%%%%%%%%%%%%%%%%%%%%%%%

At unitarity the physics is independent of the precise form of the
interaction potential. We will therefore treat the interaction among 
opposite spin particles as arising from an attractive square well of 
depth $V_0$ and range $b$. In the zero range limit only $s$-wave scattering 
contributes to the scattering amplitude. Outside the range of the potential 
the $l=0$ wave function has the form 
\begin{align}
R_{k,l=0}=\sqrt{2k^2}\,\frac{\sin(kr+\delta_0)}{kr}\; , 
\end{align}
where $\delta_0$ is the $s$-wave phase shift. The correction to the 
free $l=0$ Slater sum is given by
\begin{align}
\label{w2_del_s}
\Delta W^{(2)}_{l=0} = \frac{\lambda^3}{\sqrt{2}r^2}
  \int dk\, \left[\sin^2\left(kr+\delta_0(k)\right)-\sin^2(kr)\right]
       e^{-\frac{\lambda^2 k^2}{2\pi}}\,.
\end{align}
In principle analogous expressions can be found inside the interaction
region $r<b$.  However, at unitarity, we are interested in the limit 
$b\to 0$ while keeping $\sqrt{V_0}b=\pi/2$ fixed. We can therefore 
evaluate Eq.~(\ref{w2_del_s}) for $\delta_0=\pi/2$ independent of $k$.  
In this case the integral is straightforward. We find
\begin{align}
\Delta W^{(2)}_{l=0} 
   = \frac{2\lambda^2}{\pi r^2}e^{-2\pi r^2/\lambda^2}\; ,
\end{align}
and the effective potential at unitarity is given by
\begin{align}
\label{u_unit}
u_{eff}^{a\to \infty} 
 = -\kb T\log\left(1+\frac{\lambda^2}{\pi r^2}
        e^{-2\pi r^2/\lambda^2}\right)\; .
\end{align}
We observe that the only length scale in the potential is 
the thermal wave length. 

%%%%%%%%%%%%%%%%%%%%%%%%%%%%%%%%%%%%%%%%%%%%%%%%%%%%%%%%%%%%%%%%%%%%%%%%%
\section{Molecular dynamics simulation}
\label{sec:md}
%%%%%%%%%%%%%%%%%%%%%%%%%%%%%%%%%%%%%%%%%%%%%%%%%%%%%%%%%%%%%%%%%%%%%%%%%

Having constructed the quasi-classical effective potential we  
now describe the molecular dynamics simulations.  We consider a 
two component system with $N=N_{\uparrow} + N_{\downarrow}$ particles. 
The net polarization is zero and $N_{\uparrow}=N_{\downarrow}=N/2$.
Using Eq.~(\ref{u_id}) and (\ref{u_unit}) the two body interactions 
between like and unlike spins are\footnote{In practice one must 
soften the singular behavior of 
the attractive interaction between opposite spins. For this purpose 
we have replaced the term $1/r^2$ inside the logarithm in 
Eq.~(\ref{eq:mdpot}) by $(1+\sqrt{2}\pi\l_0)/(r^2 + \l_0^2)$, 
where $\l_0$ is a regularization parameter.  The normalization 
was chosen so that the second virial coefficient $b_2$ is insensitive 
to changes in $\l_0$. We have checked that our numerical results are
insensitive to $\l_0$ within the errors quoted in the text as long as 
$\l_0 \lsim 0.05$ in the simulation units defined below.}
\begin{align}
u_{\uparrow\downarrow} &= u_{\downarrow\uparrow} 
  = -\kb T\log\Big(1+\frac{\lambda^2}{\pi r^2}
      e^{-2\pi r^2/\lambda^2}\Big)\, , \nonumber\\
u_{\uparrow\uparrow} &= u_{\downarrow\downarrow} 
 = -\kb T\log\Big(1 - e^{-2\pi r^2/\lambda^2}\Big)\, . 
\label{eq:mdpot}
\end{align}
The molecular dynamics equations of motion are 
\begin{align}
\frac{d\vec{q}_i}{dt} = \frac{\vec{p}_i}{m}\, , \hspace{1cm}
\frac{d\vec{p}_i}{dt} = \vec{F}_i \, ,
\end{align}
where $i=1,\ldots, N$ and $F_i$ is the force on the $i$'th particle
due to the potential given in Eq.~(\ref{eq:mdpot}). We measure the 
temperature in the simulation from the average kinetic energy per 
particle, 
\begin{align}
k_B T=\frac{1}{3N}\left\langle\sum_i m\left(\frac{d}{dt}\vec{r}_i\right)^2
 \right\rangle\, , 
\end{align}
where the angular brackets denote an average over the simulation
time. The pressure is computed using the virial theorem
\begin{align}
PV = N k_B T + \frac{1}{3} 
   \left\langle \sum_i \vec{r}_{i}\cdot \vec{F}_{i}\right\rangle \, . 
\end{align}
It is advantageous to adopt a system of dimensionless units in 
which to perform the molecular dynamics simulations. We have used 
the system of units described in Table~\ref{tab:units}. In particular,
we use the thermal wave length $\lambda$ as the unit of distance, and 
$\lambda (m/T)^{1/2}$ as the unit of time. We will denote quantities
that are expressed in simulation units by a star, for example 
$r^*=r/\lambda$. In this system of units the simulation temperature 
$T^*$ is equal to unity, and the physical temperature is adjusted by 
changing the density $n^*=n\lambda^3$. The ratio $T/T_F$, where $T_F
= (3\pi^2 n)^{2/3}\hbar^2/(2mk_B)$ is the Fermi temperature, is 
given by $T/T_F = 4\pi(3\pi^2 n^*)^{-2/3}$.

%%%%%%%%%%%%%%%%%%%%%%%%%%%%%%%%%%%%%%%%%%%%%%%%%%%%%%%%%%%%%%%%%%%%%%
\begin{table*}
\begin{tabular}{l   l}
\hline
Simulation Units & \\
\hline
Mass &  $m=$ mass of one atom \\
Length &  $\lambda$ \\
Energy &  $\kb\times$system temperature \\
Time & $t^*=\lambda\sqrt{m/T}$ \\
\hline
Derived Units & \\
\hline
density & $n^* = N\lambda^3/V$ \\
Temperature & $T^* = 1$\\
Pressure & $P^* = P\lambda^3/T$ \\
Shear viscosity & $\eta^* = \eta \lambda^3/T t^*$
\end{tabular}  
\caption{System of units used in the numerical simulations of this work.}
\label{tab:units}
\end{table*}
%%%%%%%%%%%%%%%%%%%%%%%%%%%%%%%%%%%%%%%%%%%%%%%%%%%%%%%%%%%%%%%%%%%%%%%%%

 We note that the effective quasi-classical potential is temperature 
dependent. In practice we choose a simulation density $n^*=n\lambda^3$.
We initialize the simulation using a guess for the total kinetic energy.  
We monitor the kinetic energy as the system equilibrates and add or 
subtract energy by rescaling the velocities to reach the desired 
simulation temperature $T^*=1$. After the system equilibrates at this
temperature we measure observables like the equation of state, the pair 
correlation function, and the correlation function of the stress tensor. 
The main simulation parameters are listed in Table \ref{tab:params}.
Fluctuations are used to determine statistical errors. In addition, there 
are a number of systematic errors whose effect are difficult to quantify.  
One such systematic error is due to the finite number of particles.  For the 
equation of state we have performed calculations with $N=32$, $N=108$ and 
$N=256$ particles. Fig.~\ref{fig:eos} shows that the corresponding results 
agree within the statistical errors. Due to computational limitations 
we have only used runs with $N=108$ for our calculations of the pair 
and stress tensor correlation functions. Another source of systematic 
uncertainty is the fact that molecular dynamics simulations are most 
naturally performed in a microcanonical ensemble (at fixed energy). This 
requires us to tune the energy very precisely in order to reach 
the simulation temperature. This is challenging because of long  
equilibration times and finite size fluctuations.

%%%%%%%%%%%%%%%%%%%%%%%%%%%%%%%%%%%%%%%%%%%%%%%%%%%%%%%%%%%%%%%%%%%%%%
\begin{table*}
\begin{tabular}{ll}
\hline
Simulation Parameters \\
\hline
$N_{\textrm{atoms}}$ &   108\\
$l_0^*$           &  0.05 \\
$\Delta t^*$      &  0.001 \\
$t^*_{\textrm{equilibration}}$ & 2500\\
$t^*_{\textrm{thermostat}}$ & 50\\
$t^*_{\textrm{production}}$ & $2\times 10^{5}$
\end{tabular}  
\caption{Parameters used in the numerical simulations of this work.  
$N_{\textrm{atoms}}$ is the number of atoms, $l_0^*$ is the cutoff in 
the potential, $\Delta t^*$ is the molecular dynamics time step, 
$t^*_{\textrm{equilibration}}$ is the equilibration time, 
$t^*_{\textrm{thermostat}}$ is the time between velocity rescalings
during equilibration, and $t^*_{\textrm{production}}$ is the total
length of the molecular dynamics trajectory.  We have also performed 
runs with $N_{\textrm{atoms}}=32$ and $256$.}
\label{tab:params}
\end{table*}
%%%%%%%%%%%%%%%%%%%%%%%%%%%%%%%%%%%%%%%%%%%%%%%%%%%%%%%%%%%%%%%%%%%%%%%%%

%%%%%%%%%%%%%%%%%%%%%%%%%%%%%%%%%%%%%%%%%%%%%%%%%%%%%%%%%%%%%%%%%%%%%%%%%
\section{Results}
\label{sec:results}
%%%%%%%%%%%%%%%%%%%%%%%%%%%%%%%%%%%%%%%%%%%%%%%%%%%%%%%%%%%%%%%%%%%%%%%%%

%%%%%%%%%%%%%%%%%%%%%%%%%%%%%%%%%%%%%%%%%%%%%%%%%%%%%%%%%%%%%%%%%%%%%%%%%
\subsection{Equation of state}
\label{sec:eos}
%%%%%%%%%%%%%%%%%%%%%%%%%%%%%%%%%%%%%%%%%%%%%%%%%%%%%%%%%%%%%%%%%%%%%%%%%

 In this section we present our results for the equation of state. We also 
make comparisons to the available experimental data and to analytical 
results in the high temperature limit.  The pressure in the limit 
$n\lambda^3\ll 1$ is given by the virial expansion
\begin{align}
 \frac{P}{nT} \simeq 1-b_2\left(\frac{n\lambda^3}{2}\right)
  + \left(4b_2^2-2b_3\right)\left(\frac{n\lambda^3}{2}\right)^2
  + O((n\lambda^3)^3)\, .
\end{align}
The second virial coefficient is well-known \cite{Beth:1936,Ho:2004zza},
$b_2=3/(4\sqrt{2})$, and the third virial coefficient has been computed 
in \cite{PhysRevLett.98.090403,PhysRevLett.102.160401,Kaplan:2011br},
$b_3 = −0.29095295$.

\begin{figure}[t]
\begin{center}
\includegraphics[scale=.8]{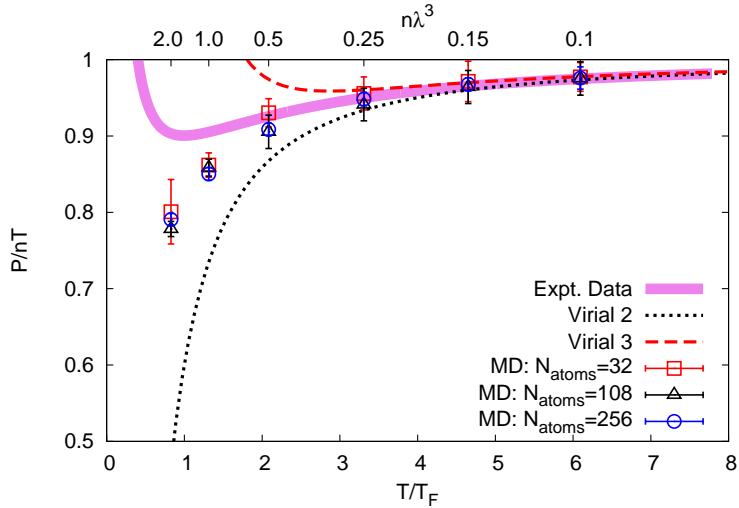}
\caption{Pressure in units of the ideal gas pressure as a
function of $T/T_F$ (bottom axis) and $n\lambda^3$ (top axis).
The data were obtained from quasi-classical molecular dynamics 
simulations with different numbers of particles in a periodic 
box. The dashed and dotted lines show the virial expansion at
second and third order, and the band is a parameterization of 
the experimental data of the ENS group \cite{nascimbene2009eos}.}
\label{fig:eos}
\end{center}
\end{figure}

In Fig.~\ref{fig:eos} we show the pressure normalized to $nT$ as a 
function of the dimensionless density $n\lambda^3$ and the temperature
in units of the Fermi temperature, $T/T_F$. The data 
points show the results of the molecular dynamics simulations. The
results are compared to the virial expansion at second and third order 
(dotted and dashed lines), and to a parameterization\footnote{See
appendix A of \cite{Schafer:2010dv}.} (band) of the experimental 
data from \cite{nascimbene2009eos}. More accurate results for the 
equation of state have recently been published by the MIT group
\cite{Ku:2011}. In the range of temperatures that are of interest
to us, $T/T_F \gtrsim 0.5$, the more recent data agrees with the 
earlier data within the errors indicated by the thickness of the 
band. 

  We observe that the data follow the second order virial expansion 
for $n\lambda^3\lsim 0.2$. This agreement is of course a consequence 
of the way the potential was constructed, but it serves as a useful 
check of the molecular dynamics (MD) simulation. We also observe that 
the full MD simulation is better behaved than the virial expansion. While 
the virial expansion is not useful for $n\lambda^3\gtrsim 0.3$ the MD 
results follow the data up to densities $n\lambda^3\sim (0.5-1.0)$.  
Assessing whether this improvement is fortuitous, or whether two-body 
interactions resummed by the MD simulation do indeed capture a 
significant part of the higher virial coefficients will require an 
explicit calculation of the three-body effective interaction, which 
we hope to pursue in a future work. The MD results do not reproduce 
the rapid increase in the pressure for $n\lambda^3\gtrsim 2$. Based 
on the sign of the third virial coefficient it is reasonable to 
assume that the MD results in this regime could be improved by 
including a repulsive three body potential.    

%%%%%%%%%%%%%%%%%%%%%%%%%%%%%%%%%%%%%%%%%%%%%%%%%%%%%%%%%%%%%%%%%%%%%%%%%
%%%%%%%%%%%%%%%%%%%%%%%%%%%%%%%%%%%%%%%%%%%%%%%%%%%%%%%%%%%%%%%%%%%%%%%%%

%%%%%%%%%%%%%%%%%%%%%%%%%%%%%%%%%%%%%%%%%%%%%%%%%%%%%%%%%%%%%%%%%%%%%%%%%
\subsection{Pair correlation function}
\label{sec:pair}
%%%%%%%%%%%%%%%%%%%%%%%%%%%%%%%%%%%%%%%%%%%%%%%%%%%%%%%%%%%%%%%%%%%%%%%%%

 For a homogeneous system the pair correlation function is defined as
\begin{align}
G(r,t) = \frac{V}{N\left(N-1\right)}\left< \sum_i \sum_{j\neq i} \delta
  \left[ r-\vert\vec{r}_i(0) - \vec{r}_j(t)\vert\; \right]\right> \, .
\label{G_def}
\end{align}
The pair correlation function measures the probability of finding 
two particles separated by a distance $r$ and time $t$. For $t=0$ the 
quantity $G(r,0)$ is also known as the radial distribution function, or as 
the Fourier transform of the static structure factor. For a two component 
system we can define two correlation functions, $G_{\uparrow\uparrow}=
G_{\downarrow\downarrow}$ and $G_{\uparrow\downarrow}=G_{\downarrow\uparrow}$,
which characterize the probability of finding two particles of the same or
opposite spin close to one another.

%%%%%%%%%%%%%%%%%%%%%%%%%%%%%%%%%%%%%%%%%%%%%%%%%%%%%%%%%%%%%%%%%%%%%%%%%
\begin{figure}[t]
\begin{center}
\includegraphics[scale=.6]{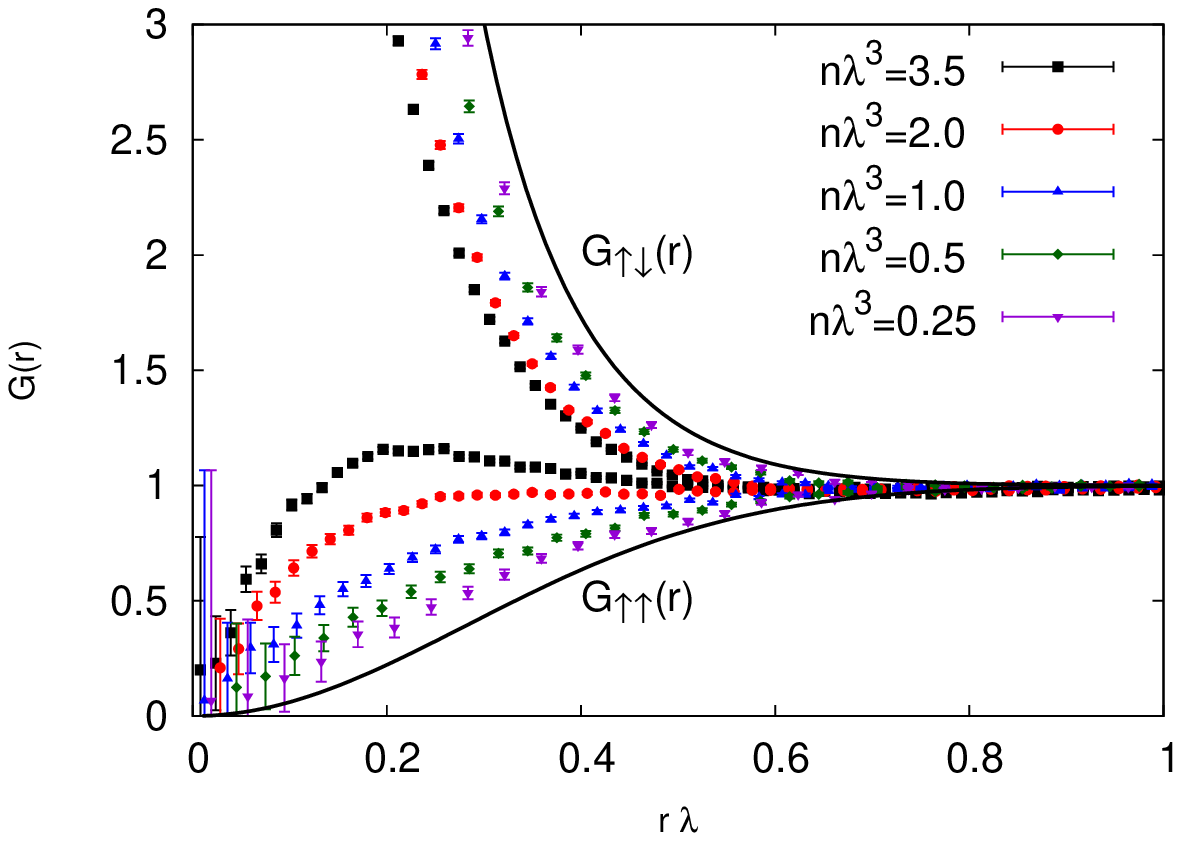}
\includegraphics[scale=.6]{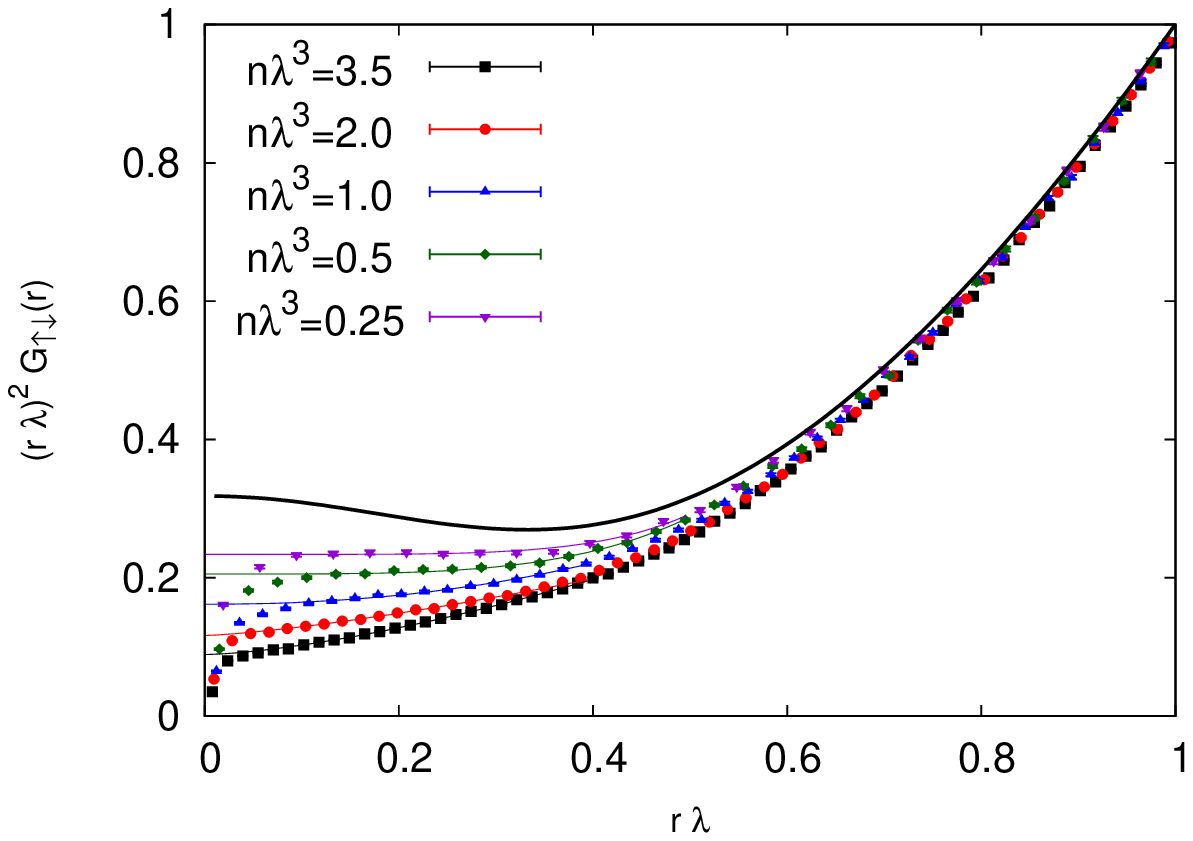}
\caption{Radial distribution functions $G_{\uparrow\uparrow}$ and
$G_{\uparrow\downarrow}$ extracted from molecular dynamics simulations
with 108 particles performed at different values of the diluteness
parameter $n\lambda^3$. The solid lines show the analytic result in
the high temperature (low density) limit. The right panel shows the
pair correlation function for unlike spins scaled by $r^2$. The 
intercept is proportional to Tan's contact density ${\cal C}$. Note
that the correlation function for $r\lambda<0.05$ is sensitive 
to the regulator in the potential. The thin solid lines show the
fit described in the text. }
\label{fig:gr}
\end{center}
\end{figure}
%%%%%%%%%%%%%%%%%%%%%%%%%%%%%%%%%%%%%%%%%%%%%%%%%%%%%%%%%%%%%%%%%%%%%%%%%

 Fig.~\ref{fig:gr} shows the radial distribution function for a system 
of 108 particles ($N_\uparrow = N_\downarrow=54$) at different densities.  
The solid curves show the high temperature ($n\lambda^3 \ll 1$) limit
\begin{align}
\label{G_ud_as}
G(r,0) = \exp\left[-\frac{u_{eff}(r)}{k_B T}\right]\, , 
\end{align} 
where $u_{eff}$ is given in Eq.~(\ref{eq:mdpot}). We observe that the 
like spin correlation function is repulsive, which is a reflection of the 
Pauli principle, and the unlike spin correlation function is strongly 
attractive, which is a consequence of the attractive interaction in 
the spin singlet channel. The range of the correlation function is equal 
to the thermal wave length. The MD results show that at non-asymptotic 
temperatures the correlation function are more short range. We also 
observe that the equal spin correlation becomes attractive at intermediate 
range, and that there is less attraction in the opposite spin channel. 
The pair correlation function at zero temperature was studied using 
Green Function Monte Carlo (GFMC) \cite{Astrakharchik:2004zz,Lobo:2006}.
At $T=0$ the pair correlation function has the same shape as in the 
high temperature limit, but the range is set by $k_F^{-1}$. 

%%%%%%%%%%%%%%%%%%%%%%%%%%%%%%%%%%%%%%%%%%%%%%%%%%%%%%%%%%%%%%%%%%%%%%%%%
\begin{figure}[t]
\begin{center}
\includegraphics[scale=.8]{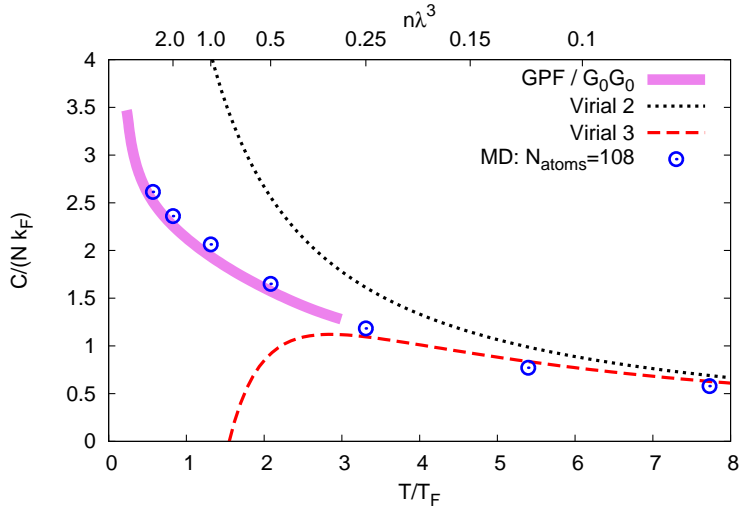}
\caption{Contact density ${\cal C}$ in units of $Nk_F$ as a function
of $n\lambda^3$ and $T/T_F$. We also show the high temperature limit
from the virial expansion at second and third order, and the result
of two T-matrix approximations, labeled GPF/G$_0$G$_0$ (see text).}
\label{fig:contact}
\end{center}
\end{figure}
%%%%%%%%%%%%%%%%%%%%%%%%%%%%%%%%%%%%%%%%%%%%%%%%%%%%%%%%%%%%%%%%%%%%%%%%%

 The right panel of Fig.~\ref{fig:gr} shows the short distance
behavior of the unlike spin correlation function. Tan observed
that the correlation function is proportional to $1/r^2$ at 
all temperatures \cite{Tan:2008a,Tan:2008b}. The $1/r^2$ term
is governed by a universal parameter known as Tan's contact
density ${\cal C}$,
\begin{align}
\label{G_ud_C}
G_{\uparrow\downarrow}(r,0)= 
   \frac{\mathcal{C}}{16\pi^2n_\downarrow n_\uparrow}
   \left(\frac{1}{r^2}-\frac{2}{ar}\right)\, . 
\end{align}
This relation combined with Eq.~(\ref{G_ud_as}) shows that the 
contact density scales as $T^{-1}$ in the high temperature limit
\cite{Yu:2009,Braaten:2010,Hu:2010},
\begin{align}
\label{C_as}
{\cal C} = \frac{32\pi^2n_\downarrow n_\uparrow}{mT}\, . 
\end{align}
Fig.~\ref{fig:gr} shows that at non-asymptotic temperatures
the contact is smaller than the limiting form given in Eq.~(\ref{C_as}).  
The pair correlation function in the limit $r\to 0$ is sensitive to the 
regulator $l_0$ in the potential. In order to extract the contact density
we fit the unlike-spin correlation functions at short distances ($l_0 \leq 
r\lambda \leq 0.5$) with the functional form
$ (r\lambda)^2 G_{\uparrow\downarrow}(r\lambda)={\cal C}/(4\pi^2n^2)
+a_0(r\lambda)^p$,
where $a_0,p$ and ${\cal C}$ are treated as fit parameters. The quality 
of the fit can be seen from the thin solid lines in the right plot of 
Fig.~\ref{fig:gr}. The value of the contact density obtained from the
fit is shown in Fig.~\ref{fig:contact}.  The size of the data points 
approximate the error in the fitted contact density estimated by varying 
the size of the fit region by $10\%$ in either direction.

 In Fig.~\ref{fig:contact} we also compare the MD result for the contact 
density to recent T-matrix calculations. The line labeled GPF/G$_0$G$_0$
shows the results for two particular truncations, the Generalized Pair 
Fluctuation (GPF) theory of Nozi\'eres and Schmitt-Rink \cite{nsr,Hu:2010}, 
and the non-self consistent (G$_0$G$_0$) T-matrix approximation of Palestini
et al.\footnote{We have taken these results from the compilation in 
\cite{Hu:2010}. The GPF and G$_0$G$_0$ results agree within the width of 
the band in Fig.~\ref{fig:contact}. The self consistent T-matrix calculation 
of Punk et al.~\cite{Punk:2007} is about 10\% higher at $T/T_F=1$. The 
recent Path Integral Monte Carlo calculation of Drut et al.~\cite{Drut:2010yn}
indicates that ${\cal C}/(Nk_F)$ reaches a maximum of $\sim 3.4$ at $T/T_F
\simeq 0.4$ and then decreases slightly to $\sim 3.0$ at 
$T=0$.}~\cite{Palestini:2010}. We find very good agreement between 
the results of the MD simulation and the T-matrix calculations for all 
densities we have studied, $0.05\lsim n\lambda^3\lsim 2$. We also note 
that the T-matrix calculations (convoluted with suitable density profiles 
for finite traps) were shown to agree with the recent data reported in 
\cite{finiteTSwinExpt}, see \cite{Hu:2010}.

%%%%%%%%%%%%%%%%%%%%%%%%%%%%%%%%%%%%%%%%%%%%%%%%%%%%%%%%%%%%%%%%%%%%%%%%%
\begin{figure}[t]
\begin{center}
\includegraphics[scale=.6]{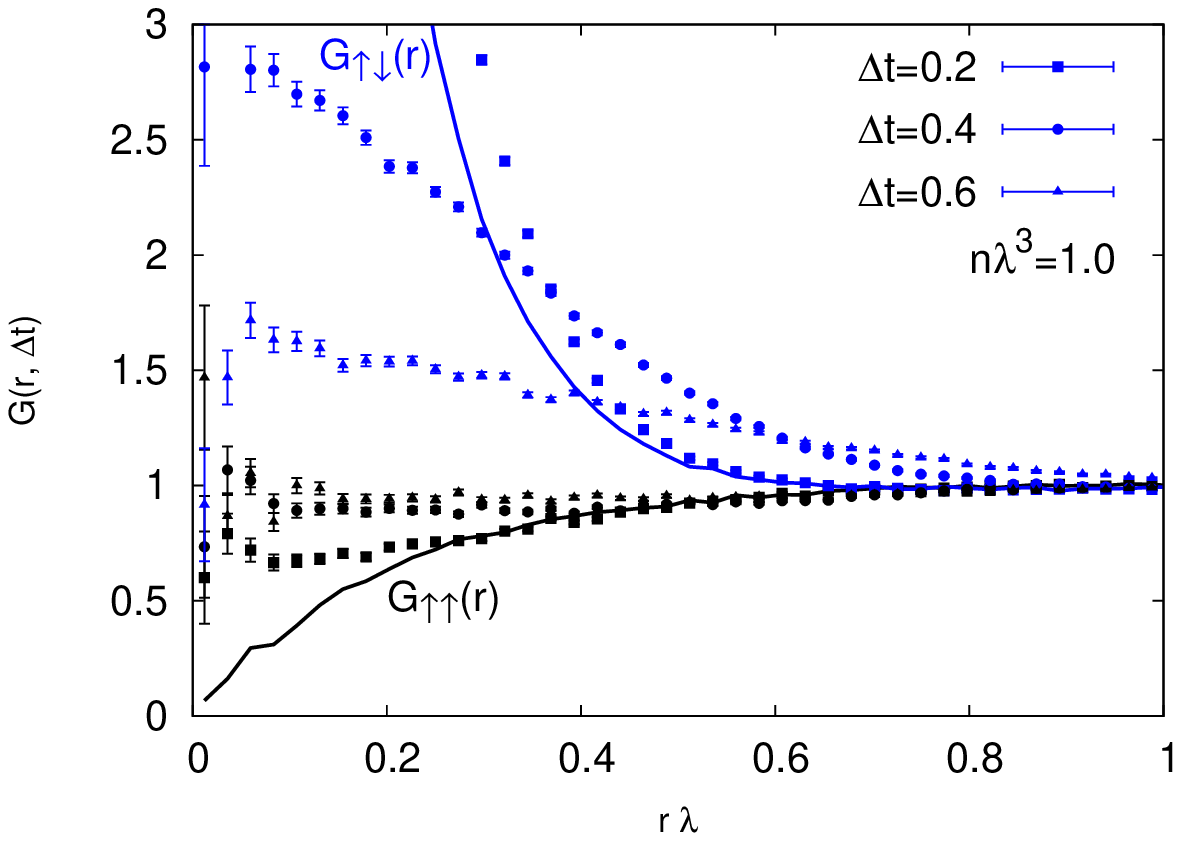}
\includegraphics[scale=.6]{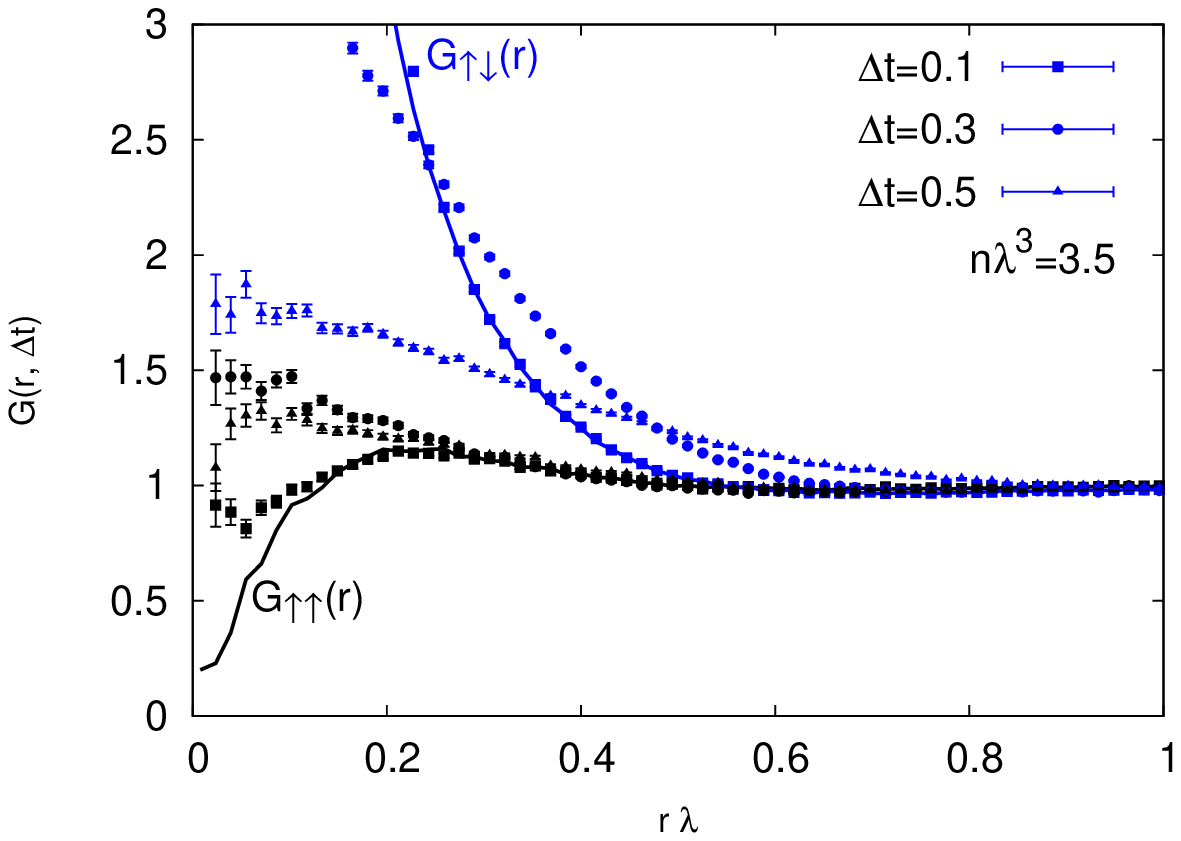}
\caption{Unequal time pair correlation function for $n\lambda^3=1.0$ 
(left) and $n\lambda^3=3.5$ (right) at $\Delta t=0.2,0.4,0.6$, where
$\Delta t$ is expressed in units of $\lambda\sqrt{m/T}$. The solid 
curves show the corresponding $\Delta t=0$ result from Fig.~\ref{fig:gr}.}
\label{fig:grtime}
\end{center}
\end{figure}
%%%%%%%%%%%%%%%%%%%%%%%%%%%%%%%%%%%%%%%%%%%%%%%%%%%%%%%%%%%%%%%%%%%%%%%%%

 Fig.~\ref{fig:grtime} shows the pair distribution function at non-zero 
time difference. This is the first quantity in this work that is 
not directly amenable to quantum Monte Carlo studies. The Fourier
transform of $G(r,t)$ yields the dynamic structure function, which 
has been studied extensively in a variety of many-body theories
\cite{bloch2008,giorgini2008}. We have not attempted to perform 
the Fourier transform, since this would require high statistics
data on a very fine mesh. We observe that the typical correlation
time in our data is of order one in simulation units $\lambda\sqrt{m/T}$.

%%%%%%%%%%%%%%%%%%%%%%%%%%%%%%%%%%%%%%%%%%%%%%%%%%%%%%%%%%%%%%%%%%%%%%%%%
\section{Shear Viscosity}
\label{sec:shear}
%%%%%%%%%%%%%%%%%%%%%%%%%%%%%%%%%%%%%%%%%%%%%%%%%%%%%%%%%%%%%%%%%%%%%%%%%

\begin{figure}[t]
\begin{center}
\includegraphics[scale=0.8]{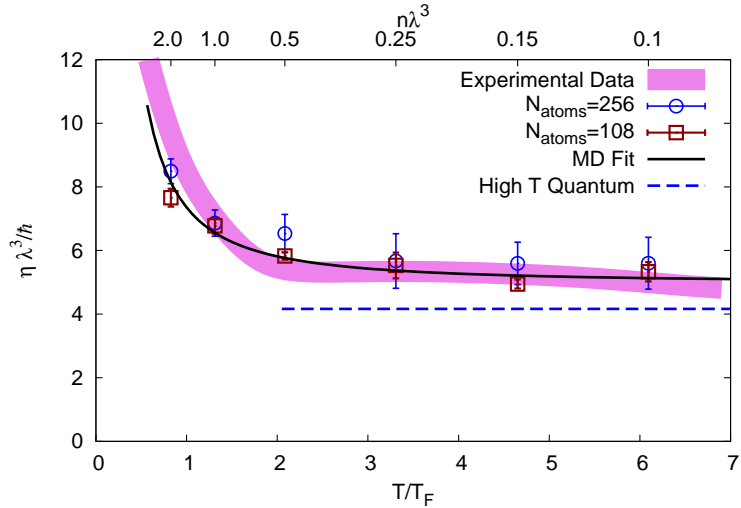}
\caption{Molecular dynamics results for the shear viscosity $\eta$
in units of $\hbar\lambda^{-3}$ as a function of $n\lambda^3$ and 
$T/T_F$. The solid black line shows the fit to the MD results as 
described in the text, and the dashed blue line shows the quantum 
mechanical high temperature result.  The band shows the experimental 
data published in \cite{Cao:2010wa,Cao:2011fh}. Note that the 
data represent trap averages, and that we have used the ideal gas
temperature at the center of the trap. }
\label{fig:etaFit}
\end{center}
\end{figure}

 We compute the shear viscosity coefficient using the Green-Kubo 
relation
\begin{align}
\eta_{(xy)} &= \frac{V}{k_B T}\int_0^\infty 
  \left\langle  P_{xy}(t)P_{xy}(0)\right\rangle dt\;,
\label{eq:kubo}
\end{align}
where $P_{xy}(t)$ is the stress energy tensor evaluated at time $t$ 
\cite{Haile:1992:MDS:531139,Rapaport:2004},
\begin{align}
P_{xy}(t) = m\sum_{i=1}^N \dot{x}_i\dot{y}_i 
  + \frac{1}{2} \sum_{i=1}^N\sum_{j\neq i}^N 
  \frac{1}{r}\frac{\partial u_{ij}}{\partial r}
   \left(x_i-x_j\right)\left(y_i-y_j\right)\;.
\end{align}
There are five independent stress tensor correlation functions that can 
be used to determine the shear viscosity. We denote the corresponding 
estimates by $\eta_{(xy)},\eta_{(yz)},\eta_{(xz)},\eta_{(xxyy)}$ and 
$\eta_{(yyzz)}$, and use the average of the five measurements as our final 
result\footnote{$\eta_{(yz)}$ and $\eta_{(xz)}$ can be obtained trivially 
from Eq.~(\ref{eq:kubo}). The diagonal $(xxyy)$ Kubo formula is given by 
$\eta_{(xxyy)}=\frac{V}{k_B T}\int_0^\infty \left\langle\left(P_{xx}(t)-P_{yy}(t)
\right)\left(P_{xx}(0)-P_{yy}(0)\right)\right\rangle dt$, and $\eta_{(yyzz)}$ 
is defined analogously.}. The results for $N=108$ and $N=256$ is shown 
in Fig.~\ref{fig:etaFit}.

 We can check the MD simulation by comparing the numerical result to 
the expected behavior in the high temperature limit. The calculation 
of the transport cross section and the shear viscosity is explained 
in the Appendix. For a two component system with the classical interaction 
given in Eq.~(\ref{eq:mdpot}) the high temperature limit of the shear 
viscosity is
\begin{align}
\left.\frac{\eta}{\hbar n}\right|_{\textrm{cl}}
=\frac{75\sqrt{2}\pi}{8\left(5+\pi\right)n\lambda^3}\; .
\label{eq:vis2compCl}
\end{align}
This result is shown as the high temperature limit of the solid 
black line label `MD Fit' in Fig.~\ref{fig:etaFit}, and we observe that the agreement 
with the MD results in the regime $n\lambda^3 \lesssim 0.15$ is very good. 
The classical result is about 20\% larger than the (almost exact) quantum 
result in the high temperature limit 
\cite{Massignan:2005zz,Bruun:2006kj,Bruun:2005en}  
\begin{align}
\frac{\eta}{\hbar n}
 = \frac{45\pi^{3/2}}{64\sqrt{2}}\left(\frac{T}{T_F}\right)^{3/2}
 = \frac{15\sqrt{2}\pi}{16(n\lambda^3)}\; .
\end{align}
This result is shown as the dashed blue line in Fig.~\ref{fig:etaFit}.
The discrepancy between the classical and quantum calculation is due
to a combination of two effects, discussed in more detail in 
Appendix~\ref{app:chapman}. The first is the fact that the unlike 
spin potential does not exactly reproduce the quantum mechanical 
transport cross section. The second effect is that the like spin
potential, related to Pauli repulsion, leads to a finite transport 
cross section, even though there is no scattering in a quantum 
system of like spins interacting by a pure $s$-wave potential.

%%%%%%%%%%%%%%%%%%%%%%%%%%%%%%%%%%%%%%%%%%%%%%%%%%%%%%%%%%%%%%%%%%%%%%%%%
%%%%%%%%%%%%%%%%%%%%%%%%%%%%%%%%%%%%%%%%%%%%%%%%%%%%%%%%%%%%%%%%%%%%%%%%%

 Fig.~\ref{fig:etaFit} also shows a parameterization of the viscosity 
measured in \cite{Cao:2010wa,Cao:2011fh}.  The thickness of the band 
approximates the statistical errors in the measurement.  Note that the 
measured viscosity is a trap averaged quantity, and as a result 
there is a $\sim 20\%$ discrepancy between the data and the theoretical
result in the high temperature limit (dashed line). The very close
agreement between the measured viscosity and the MD simulations is 
therefore somewhat of a coincidence. 

 We note that the molecular dynamics simulation reproduces the $\lambda^{-3}$ 
scaling of the shear viscosity. We also emphasize that the numerical 
discrepancy between the classical and quantum result is quite small.
It is therefore reasonable to extract an estimate of the leading 
density dependence of the shear viscosity from the MD simulation. 
We have fit the MD results shown in Fig.~\ref{fig:etaFit} with the 
expression
\begin{align}
\eta = \frac{\eta_0}{\lambda^3} 
   \left( 1 + c_2 \left(n\lambda^3\right) \right) \; , 
\end{align}
where $\eta_0$ was fixed using Eq.~(\ref{eq:vis2compCl}). We find
$c_2\simeq 0.32$.  The fit well represents the MD results in the
density range studied in this work, $0.1\leq (n\lambda)^3\leq 2$, including the rise in 
$\eta/\lambda^3$ around $(n\lambda)^3\sim (0.5-2.0)$, which is also seen in
the experimental data.

%%%%%%%%%%%%%%%%%%%%%%%%%%%%%%%%%%%%%%%%%%%%%%%%%%%%%%%%%%%%%%%%%%%%%%%%%
\section{Discussion and future work}
\label{sec:out}
%%%%%%%%%%%%%%%%%%%%%%%%%%%%%%%%%%%%%%%%%%%%%%%%%%%%%%%%%%%%%%%%%%%%%%%%%

 We have presented a new method for studying the dynamics of cold 
atomic gases based on molecular dynamics simulations with an effective 
quantum potential. In this work we have restricted ourselves to two-body 
interactions. In this case the MD simulation exactly reproduces the 
second virial coefficient and the pair correlation function in the 
dilute limit. We have measured the equation of state, the pair 
correlation function, and the shear viscosity for a range of densities 
$0.1\lesssim n\lambda^3\lesssim 2.0$. We find that we can reproduce the 
experimentally measured equation of state for densities $n\lambda^3 
\lesssim (0.5-1.0)$. This is an improvement over the virial expansion, 
which breaks down $n\lambda^3 \gtrsim 0.3$, and we expect that the 
range of applicability can be extended by including a three-body force. 

 We have also measured the static and dynamic pair correlation functions
as well as the shear viscosity. We find that in the dilute limit the
contact ${\cal C}$ scales as $n_\uparrow n_\downarrow/(mT)$, in agreement 
with the prediction in \cite{Yu:2009,Hu:2010}. Higher order correlations 
suppress the contact relative to the asymptotic behavior.  Excellent 
agreement between the MD simulations and T-matrix calculations is seen 
for $n\lambda^3 \lesssim 2$.

 In this regime the temperature dependence of the shear viscosity 
is well described by the functional form $\eta = \eta_0\lambda^{-3} 
( 1 + c_2  n\lambda^3 )$. The scaling of $\eta$ with $\lambda^{-3}$ 
agrees with the expected behavior at unitarity. The parameter $\eta_0$ 
differs from the exact quantum mechanical behavior at high temperature 
by a factor 1.2. We discuss the origin of this factor in the appendix,
but it would clearly be desirable to understand the physical origin of 
the discrepancy more clearly, and to investigate whether there are 
any improvements of the quasiclassical MD method that reproduce the 
correct transport cross section. The main result of the MD simulation
is that the density dependence of the shear viscosity is weak, $c_2 
\simeq 0.32$, and that it tends to increase the shear viscosity.

 There are a number of additional applications or further
extensions of the work presented here:

\begin{enumerate}

\item BEC-BCS crossover: In this work we focused on a dilute Fermi
gas at unitarity.  Clearly, the same methods can be used to determine 
the effective potential and the leading $n\lambda^3$ correction to 
pair correlation functions and transport coefficients for the full 
BEC-BCS crossover.

\item Three-body forces: The method discussed in this work can be 
systematically improved by including $n$-body forces ($n\geq 3$). 
In the case of three-body forces this appears quite tractable. 
The three-particle Slater sum can be computed in hyper-spherical
coordinates, and molecular dynamics with three-body interactions is
computationally feasible.

\item Other transport coefficients: The methods used in this work
can also be used to measure other transport properties like 
the spin diffusion constant studied theoretically in \cite{Bruun:2011}
and recently measured by the MIT group \cite{sommer2011uni}, or 
the thermal conductivity \cite{Braby:2010ec}. It would be 
interesting to determined whether $n\lambda^3$ corrections are
universal. It is also interesting to determine the bulk viscosity
as one goes away from the unitary limit $a\to \infty$. At unitarity
the bulk viscosity is expected to vanish based on theoretical 
reasons \cite{Son:2005tj}, and existing experiments are consistent
with these arguments \cite{Dusling:2011dq}.

\item Lower dimensional systems: Recently Vogt et al.~measured
the shear viscosity of a two-dimensional Fermi gas \cite{Vogt:2011}.
In two dimensions fluctuations are expected to be more important 
than in three dimensional systems. These effects are difficult 
to include in kinetic theory, but are automatically included 
in molecular dynamics.

\end{enumerate}

\section*{Acknowledgments}
We would like to thank Cliff Chafin, Dean Lee, Lubos Mitas and John Thomas for enlightening
discussions.  This work was supported by the US Department
of Energy Grant No. DE-FG02-03ER41260.

%%%%%%%%%%%%%%%%%%%%%%%%%%%%%%%%%%%%%%%%%%%%%%%%%%%%%%%%%%%%%%%%%%%%%%%%%
\appendix
\section{Shear viscosity of dilute classical and quantum gases}
\label{app:chapman}
%%%%%%%%%%%%%%%%%%%%%%%%%%%%%%%%%%%%%%%%%%%%%%%%%%%%%%%%%%%%%%%%%%%%%%%%%

 In this appendix we summarize the calculation of the shear viscosity 
for a dilute classical gas interacting via the potentials given in
Eq.~(\ref{eq:mdpot}). This result provides an important check for
the calculation of the viscosity using the Green-Kubo formula. For 
both quantum and classical gases the calculation of the shear 
viscosity using the Chapman-Enskog method \cite{ChapmanCowling} yields
\begin{align}
\eta = \frac{5}{8} \frac{\sqrt{\pi m k_B T}}{\sigbar}\, ,
\label{eq:shearsigma}
\end{align}
where $\sigbar$ is the energy averaged transport cross section
\begin{align}
\sigbar = \int_0^\infty \gamma^7 e^{-\gamma^2} \;\sig\; d\gamma\, , 
\end{align}
and $\sig$ is the transport cross section. The parameter $\gamma=
\sqrt{E/k_B T}$ is the dimensionless relative velocity of the 
two-particle system.  The calculation of the transport cross section 
is different in the classical and quantum case, both because the cross 
section is represented in a different way, and because of the presence of 
symmetry factors in the quantum mechanical calculation.

%%%%%%%%%%%%%%%%%%%%%%%%%%%%%%%%%%%%%%%%%%%%%%%%%%%%%%%%%%%%%%%%%%%%%%%%%
\subsection{Dilute classical gas}
%%%%%%%%%%%%%%%%%%%%%%%%%%%%%%%%%%%%%%%%%%%%%%%%%%%%%%%%%%%%%%%%%%%%%%%%%

 In the classical case the cross section is computed from classical
trajectories in the potential. We can write $\sig$ as a one dimensional 
integral over the impact parameter $b$
\begin{align}
\sig = 2\pi \int_0^\infty \left[1-\cos^2\chi\left(b,E\right)\right] b\; db\, . 
\end{align}
In this expression the scattering angle $\chi(b,E)$ is a function of 
the impact parameter and of the kinetic energy in the center-of-mass 
system through the relation
\begin{align}
\chi(b,E) = \pi-2b\int_{r_0(b,E)}^\infty \frac{r^2}{1-b^2/r^2- U(r)/E}\, , 
\end{align}
where $U(r)$ is the interaction potential and $r_0$ is the distance of 
closest approach. The parameter $r_0$ is determined by the solution of
\begin{align}
1-b^2/r_0^2-U(r_0)/E=0 \; .
\end{align}
For the potentials of interest, Eq.~(\ref{eq:mdpot}), all of the above
expressions must be evaluated numerically.  We find the following results 
for a two particles interacting through the $u_{\uparrow\downarrow}$  and 
$u_{\uparrow\uparrow}$ potentials,
\begin{align}
\sigbar^{\uparrow\downarrow}&= \frac{2}{3} \lambda^2\times (1.004)\, ,
\label{eq:classcross}\\
\sigbar^{\uparrow\uparrow}  &= \frac{2\pi}{15}\lambda^2\times \, (1.0006)\, ,
\end{align}
where the factors 1.004 and 1.006 are the results of numerical integrals. 
This work studies a two-component gas with equal numbers of 
spin up and down particles. In the dilute limit this system can be treated 
as a classical mixture and the effective transport cross section is the 
average of the $\sigma_\textrm{tr}^{\uparrow\downarrow}$ and 
$\sigma_\textrm{tr}^{\uparrow\uparrow}$ cross sections.  The shear viscosity is
\begin{align}
\left.\frac{\eta}{\hbar n}\right|_{\textrm{cl}}
=\frac{75\sqrt{2}\pi}{8\left(5+\pi\right)n\lambda^3}\, , 
\label{eq:clmix}
\end{align}
which agrees with the high temperature MD simulation as shown in 
Fig.~\ref{fig:etaFit}.

%%%%%%%%%%%%%%%%%%%%%%%%%%%%%%%%%%%%%%%%%%%%%%%%%%%%%%%%%%%%%%%%%%%%%%%%%
\subsection{Dilute quantum gas}
%%%%%%%%%%%%%%%%%%%%%%%%%%%%%%%%%%%%%%%%%%%%%%%%%%%%%%%%%%%%%%%%%%%%%%%%%

 The quantum mechanical cross section can be represented in terms
of the scattering phase shifts. In quantum mechanics we also have
to take into account the symmetry of the wave function, which depends 
on whether the particles are distinguishable or not. For distinguishable 
particles the transport cross section is \cite{ChapmanCowling}
\begin{align}
\sig = \frac{4\pi}{k^2} \sum_l \frac{(l+1)(l+2)}{(2l+3)}
      \sin^2\left[\delta_{l+2}(k)-\delta_l(k)\right]\, , 
\end{align}
where $k\equiv\sqrt{mE}/\hbar$ with $E$ the center of mass energy. 
The parameter $\gamma$ is given by $\gamma\equiv\sqrt{E/k_B T}$ as
above. For indistinguishable particles 
\begin{align}
\sig = \frac{8\pi}{k^2}\sum_{l=\textrm{e/o}}  \frac{(l+1)(l+2)}{(2l+3)}
   \sin^2\left[\delta_{l+2}(k)-\delta_l(k)\right]\, ,
\end{align}
where the sum is restricted to even/odd (e/o) angular momenta for
bosons/fermions. For a two-component Fermi gas interacting via
an $s$-wave interaction we can treat the collisions between atoms
of opposite spin as occurring between distinguishable particles. 
At unitarity the transport cross section is 
\begin{align}
\sigma^{\uparrow\downarrow}_{\textrm{tr}}(\gamma) = \frac{8\pi}{3k^2} \, ,
\end{align}
and the energy averaged cross section is given by
\begin{align}
\sigbar^{\uparrow\downarrow}=\frac{4}{3}\lambda^2 \; .
\end{align}
We find that the quantum cross-section is a factor of two larger than 
the analogous classical cross-section, Eq.~(\ref{eq:classcross}).
As in the classical case, the effective transport cross-section entering 
into the viscosity consists of an average over the $\sigma^{\uparrow\downarrow}$ 
and $\sigma^{\uparrow\uparrow}$ channels.  As the latter is zero the overall 
cross-section is reduced by a factor of two.  The final result is 
\begin{align}
\frac{\eta}{\hbar n} =\frac{15\sqrt{2}\pi}{16(n\lambda^3)}\;.
\label{eq:qumix}
\end{align}

%%%%%%%%%%%%%%%%%%%%%%%%%%%%%%%%%%%%%%%%%%%%%%%%%%%%%%%%%%%%%%%%%%%%%%%%%

\end{document}